\documentclass{cimento}
\usepackage{graphicx}

\title{Open issues in gamma-ray bursts: polarimetry and dark GRBs}

\author{
  D. Malesani\from{i1},
  S. Covino\from{i2},
  E.M. Rossi\from{i3},
  D. Lazzati\from{i4},
  A. De Luca\from{i5},
  P. Filliatre\from{i6}\from{i7} \atque
  G. Tagliaferri\from{i2}
}

\instlist{
  \inst{i1} International school for advanced studies (SISSA), via Beirut 2-4, I-34014 Trieste, Italy
  \inst{i2} INAF/Osservatorio Astronomico di Brera, via E. Bianchi 46, I-23807, Merate (LC), Italy
  \inst{i3} Max Planck Institut f\"ur Astrophys., Karl-Schwarzschild-Str. 1, 85741 Garching, Germany
  \inst{i4} JILA, 440 UCB, University of Colorado, Boulder, CO, 80309-0440, USA
  \inst{i5} INAF/IASF, sezione di Milano ``G. Occhialini'', via Bassini 15, I-20133 Milano, Italy
  \inst{i6} Laboratoire Astroparticule et Cosmologie, UMR 7164, 11 place Marcelin Berthelot, F-75213 Paris Cedex 05, France
  \inst{i7} Service d'Astrophysique, CEA/DSM/DAPNIA/SAp, CE-Saclay, Orme des Merisiers, B\^{a}t. 709, F-91191 Gif-syr-Yvette Cedex, France
 }

\PACSes{}

\begin{document}

\maketitle

\PACSes{
\PACSit{98.70.Rz}{gamma-ray surces; gamma-ray bursts},
\PACSit{95.30.Gv}{Radiation mechanisms; polarization}.
}

\begin{abstract}
  We review some open problems in the physics of afterglows, namely
  their polarization properties and the existence of dark/faint
  bursts. Polarization studies yield precious insights in the physical
  structure and dynamical evolution of GRB jets, revealing their
  magnetization properties and their energy profile. Polarimetric
  observations of GRB\,020813 already allowed to exclude a homogeneous
  jet for this event. We then present observations of faint/dark bursts,
  showing that some of them may be obscured by dust, while others are
  possibly just intrinsically dim.
\end{abstract}

\section{Polarization}

In 1999 the first successful detection of optical polarization from a
gamma-ray burst (GRB) afterglow was obtained
\cite{ref:Cv99,ref:Wij99}. Since then, several polarimetric measurements
of GRB afterglows have been performed (see \cite{ref:Cov04} and
\cite{ref:Bjor03} for a review). Albeit being an observationally
challenging task, the scientific community has shown wide interest in
this field. Indeed, polarization has a high diagnostic power over a
broad range of physical processes, from the emission mechanisms to the
fireball structure and the properties of the close and far environment
of the burst.

The search for polarizion was driven by the hypothesis that the
afterglow emission is due to synchrotron emission
\cite{ref:Pac93,ref:Mesz97,ref:Sari98}. Its discovery, by itself, is
now regarded as a strong evidence for this emission mechanism. For
unresolved sources, like distant GRB afterglows, polarimetry also offers
a unique opportunity to probe the geometry of the system. In fact, in
order to have a net nonzero polarization, some kind of asymmetry is
required, provided for example by a collimated fireball. Time-resolved
polarimetry is also a reliable tool to discriminate among different
scenarios for the blastwave evolution. Last, polarimetry of GRB
afterglows also offers a direct way to study the interstellar medium
around GRB progenitors and in general along the line of sight.

\begin{figure}
  \centering\includegraphics[width=0.70\textwidth]{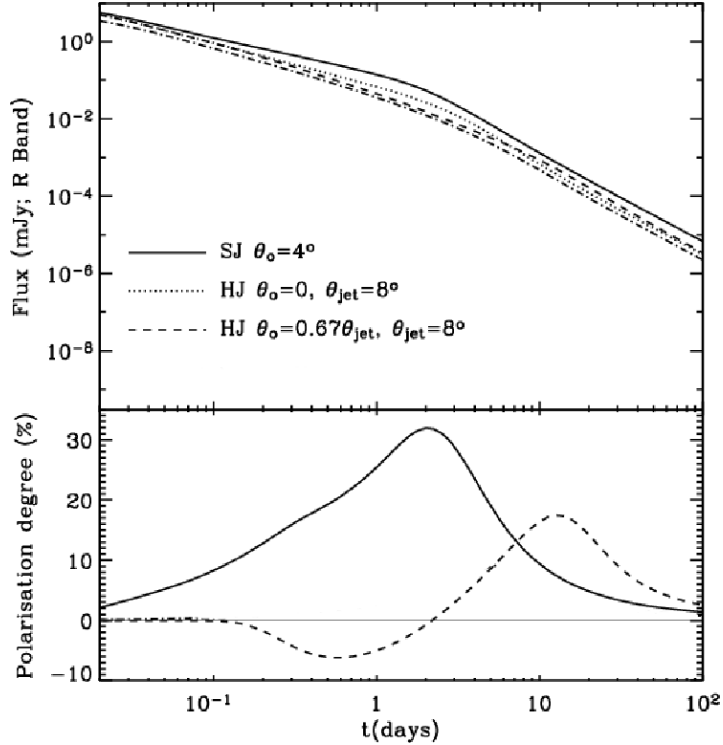}
  \caption{Light curve and polarization evolution for different jet
  structures. SJ stands for structured jet, HJ for homogeneous jet, GJ
  for Gaussian jet. The figure shows the similarity of the predicted
  light curves for the various models, in contrast with the considerable
  differences in the polarization curves. Negative polarization degrees
  indicate a 90$^\circ$ rotation for the position angle. From
  \cite{ref:Ros04}.\label{fig:polev}}
\end{figure}

The diagnostic power of polarimetry rests upon the characteristic time
evolution of the polarization degree and position angle
\cite{ref:Ghis99,ref:Sar99} produced by an ultrarelativistic
outflow. In the simplest case, the angular energy distribution is
homogeneous. More physical models consider more complex beam and
magnetic field patterns \cite{ref:Gran03,ref:Laz04,ref:Ros04}. These
works show that, even if the light curve is barely affected by these
parameters, the polarization and position angle evolutions change
substantially (fig.~\ref{fig:polev}). In the homogeneous jet (HJ) model,
the polarization curve has two maxima bracketting the jet break, and,
more important, the polarization angle has a sudden rotation of
90$^\circ$ at the same moment (fig.~\ref{fig:polev}, dashed line). On
the contrary, the structured jet (SJ) model predicts that the maximum of
the polarization curve is reached right at the time of the break in the
light curve, the position angle keeping constant throughout the
afterglow evolution (fig.~\ref{fig:polev}, solid line). At early and
late time the polarization should essentially vanish in either cases.

This latter fact constitutes an important point because, indipendently
of the model details, the prediction for the early-time polarization
changes substantially if the fireball expansion is driven by a
large-scale magnetic field. This important issue has been recently
developed and discussed, e.g., by \cite{ref:Gran03}, \cite{ref:Laz04}
and \cite{ref:Ros04}. Like hydrodynamic jets, magnetized ones can be
homogeneous and structured. In any case, a non negligible degree of
polarization at early times is expected, a strong difference with
respect to purely hydrodynamical models. Polarimetry may therefore be
the most powerful available tool to investigate the fireball structure
and its early dynamical evolution.

\begin{figure}
  \centering \includegraphics[width=0.8\textwidth]{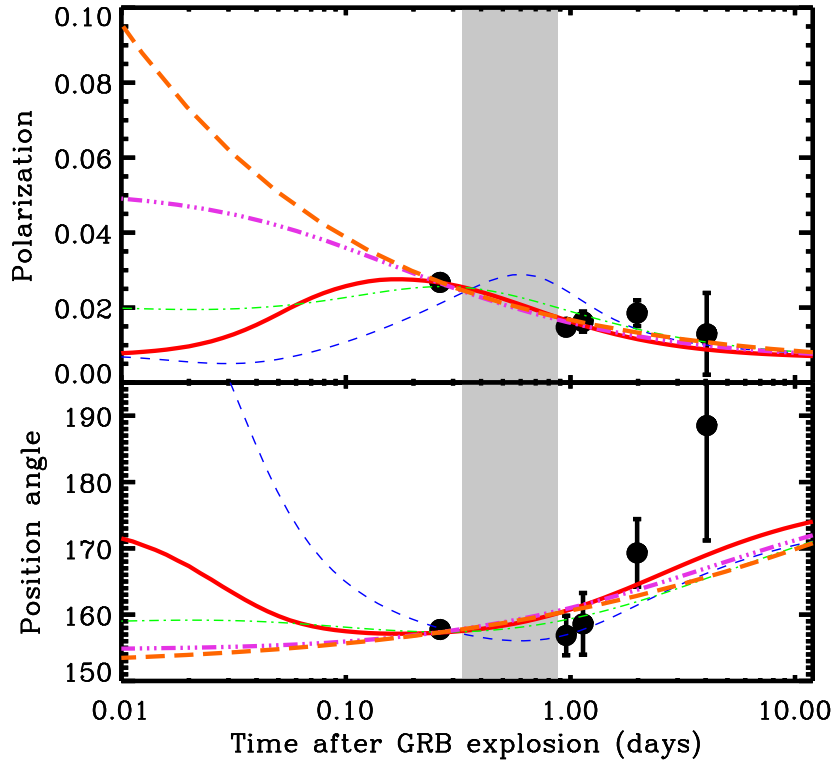}
  \caption{Polarization data for GRB\,020813 \cite{ref:Goro04}.
  Different curves refer to different models. The gray lane indicates
  the position of the jet break as measured from the light curve. From
  \cite{ref:Laz04}.\label{fig:020813}}
\end{figure}

From the observational point of view, besides several isolated
measurements, a rich enough coverage of the polarization evolution could
be obtained only in three cases: GRB\,020813
\cite{ref:Goro04,ref:Laz04}, GRB\,021004
\cite{ref:Rol03,ref:Laz03,ref:Nak04,ref:Bjor04}, and GRB\,030329
\cite{ref:Gre03,ref:Klo04}. However, firm conclusions could be derived
only for GRB\,020813, the best available case for model testing. Its
light curve was remarkably smooth \cite{ref:Cov03,ref:LauSta} and a
break in the light curve could be clearly singled out. A bunch of
polarimetric observations were carried out providing for the first time
polarization data both before and after the light curve break time
\cite{ref:Goro04}. \cite{ref:Laz04}, with a quantitative
approach, carried out a formal analysis by taking into account both the
dust-induced (host galaxy + Milky Way) and the intrinsic afterglow
polarization. All current jet models were considered, including
homogeneous and structured jets, with and without a coherent magnetic
field. The available dataset did not allow us to single out a unique
best-fitting model. However, it was possible to rule out homogeneous jet
models at a confidence level larger than $3\sigma$, mainly due to the
lack of the predicted $90^\circ$ position angle rotation. This is an
important result with possible consequences for the use of GRBs as
probes for cosmology structure studies. All magnetized models and
structured jets fit satisfactorily the data, the ambiguity being mainly
due to the lack of early-time measurements, i.e. when the magnetization
properties mostly matter (fig.~\ref{fig:020813}).

\section{Dark GRBs}

Since the beginning of afterglow observations, it was apparent that a
fraction of GRBs (the so-called dark GRBs) did not show any detectable
optical counterpart, the first example being GRB\,970828
\cite{ref:Dj01}. Reaction times and sensitivity have constantly
improved, allowing faster and deeper searches, and the dark GRB fraction
has continuously decreased. Also, the definition of ``darkness'' is
subject to discussion. Likely, a good classification should make use of
information from other bands (especially the X rays;
\cite{ref:Jako04,ref:Rol05}), in order to individuate anomalies in the
spectral shape and single out the events with an optical deficit. The
X-ray radiation is in fact much less affected by extinction, and does
not suffer from Ly$\alpha$ suppression at large redshifts. Moreover,
from the observational point of view, an X-ray afterglow was discovered
in virtually all cases.

Several solutions are viable to explain the existence of dark bursts,
and probably more than one mechanism is at work. The simplest idea is
that dark GRBs have just very faint afterglows
\cite{ref:Fynbo01,ref:Berg02}, due perhaps to the different conditions in
the environment or in the emission mechanism. This hypothesis is
supported by the analysis of BeppoSAX X-ray data
\cite{ref:DePa03}. Another possibility is that these afterglows are
heavily extinguished by material along the line of sight
\cite{ref:Laz02,ref:Rei02}, a likely possibility given the association
between GRBs and young stars \cite{ref:Stan03,ref:Hjo03}. Last, an
intriguing possibility is that some GRBs are optically dark since they
are at high redshift ($z > 7$), so that the Ly$\alpha$ dropout
suppresses visible radiation. In the last two cases, observing in the
near infrared (NIR) alleviates the problem, since this band is less
affected by dust and dropout (up to $z = 20$) extinction.

To tackle this issue, our group has undertaken an observing campaign
devoted to detect and follow-up GRB afterglows both in the optical and
in the NIR, in order to spot dark/faint/extinguished events. We report
here about three INTEGRAL bursts.

\begin{figure}\centering
  \includegraphics[width=0.8\textwidth,keepaspectratio]{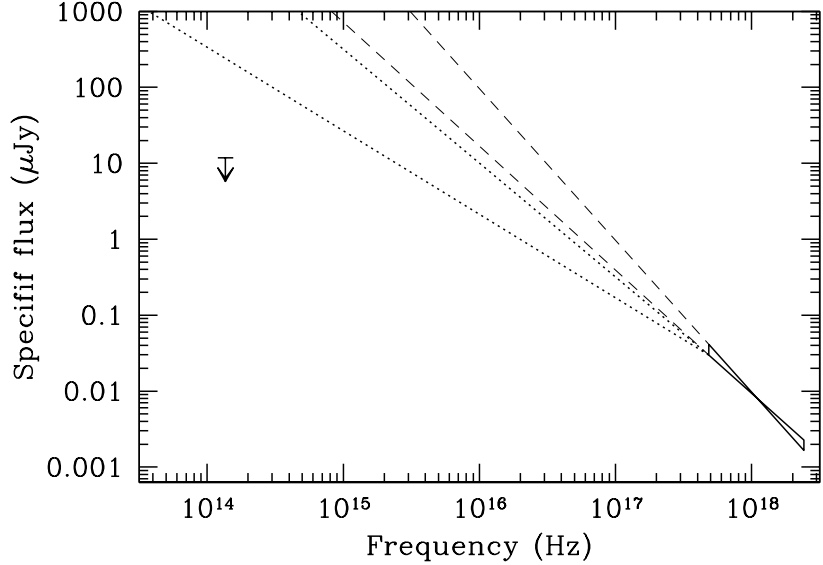}
  \caption{Broad-band spectral energy distribution of the afterglow of
    GRB\,040223, 0.7~days after the GRB. See text for the meaning of the
    lines.\label{fig:SED}}
\end{figure}

\paragraph{GRB\,040223} The long-duration GRB\,040223 had a peak flux of
$3 \times 10^{-8}$~erg~cm$^{-2}$~s$^{-1}$ ($20 \div 200$~keV;
\cite{ref:Gotz04}). Following the discovery of the X-ray afterglow by
XMM-Newton \cite{ref:Breit,ref:Gon04,ref:Tien04}, we observed the field
with the ESO-NTT telescope at several epochs, in the $JHK_\mathrm{s}$
filters. Despite our images are quite deep, no variable NIR afterglow
was detected \cite{ref:Simo04}. Figure~\ref{fig:SED} shows the
NIR-to-X-ray spectral energy distribution, showing that the
$K_\mathrm{s}$ datum lies well below the X-ray extrapolation (dashed
lines). Also, the presence of a break (e.g. due to the cooling
frequency) cannot explain the NIR faintness: given the X-ray spectral
index $\alpha_X = 1.8 \pm 0.2$ \cite{ref:Tien04}, in the standard
syncrotron model the slope redward of the break should be $\alpha =
\alpha_X - 0.5 = 1.3 \pm 0.2$ (dotted lines). Even assuming that this
break is just at the edge of the observed X-ray band, the NIR point is
still a factor at least $\sim 20$ below the extrapolation. Given the
large measured column density, $N_\mathrm{H} = (1.75 \pm 0.20) \times
10^{22}$~cm$^{-2}$, well in excess with respect to the Galactic value,
we therefore conclude that this burst is likely significantly
extinguished (S. Covino et al., manuscript in preparation).

\paragraph{GRB\,040422} This faint burst was observed by INTEGRAL,
lasting 8 s \cite{ref:Mere04}. VLT observations started very soon, just
2 hours after the GRB, in the $R$, $I$, and $K_\mathrm{s}$ bands. In
this case, despite the large extinction and the field crowding, an
afterglow could be discovered in the $K_\mathrm{s}$ band
\cite{ref:Fill05}. This is one of the faintest objects ever detected,
and only by promptly reacting with a large telescope could the afterglow
be discovered. A redshift is lacking for this event, thereby the
energetics is unknown. However, a bright host galaxy ($K_\mathrm{s} =
20.3 \pm 0.2$, one of the brightest among GRB hosts) was discovered
coincident with the afterglow, suggesting a closeby event. Also, the
afterglow was quite faint when compared to its host galaxy (being
2.3~mag brighter 2~h after the GRB). All these facts suggest a very
faint event, perhaps bridging classical GRBs and dim events like
GRB\,980425 and GRB\,031203 (the ``w's'').

\paragraph{GRB\,040827} This burst was also discovered by INTEGRAL
\cite{ref:Mere04b}, showing no remarkable properties in its gamma-ray
emission. Unlike most INTEGRAL triggers, it was not on the Galactic
plane, allowing more effective observations. In this case, an X-ray
afterglow was discovered \cite{ref:Rod04}, in turn allowing the
discovery of a NIR transient by several groups
\cite{ref:KapBe04,ref:Tan04,ref:Male04}. Also in this case, the afterglow
was quite faint when compared with the host galaxy (no redshift is
available), suggesting an intrinsically faint event. Analysis of X-ray
data \cite{ref:DeLu05} showed a significant extinction ($N_\mathrm{H}
\sim 10^{22}$~cm$^{-2}$, somewhat uncertain due to the unknown
redshift), well in excess with respect to the Galactic value. This case
may indeed be the best example of an extiguished GRB. Optical limits are
consistent with this column.\bigskip

Observation of a large number of bursts is now possible, thanks to Swift
\cite{ref:Gehr04}. Many Swift afterglows are indeed among the faintest
ever observed, promising to have soon a large sample. Among these
events, some will be just ``faint'', and some will be
extinguished. Coupling optical and X-ray observations will allow to
clearly disentangle this issue, and to select really dark bursts,
possibly at very high redshift.

\end{document}